\documentclass[a4paper]{jpconf}
\usepackage{graphicx}
\usepackage{amsmath}
\usepackage{amssymb}
\usepackage{wrapfig}
\usepackage{soul}

\usepackage{color}

\newcommand{\eO}{\vec{e}_{\Omega}}

\begin{document}
  \title{Precession and glitches in the framework of three-component model of neutron star}
  \author{
  D~P~Barsukov$^{1,2}$,  
  O~A~Goglichidze$^{2}$,
  K~Y~Kraav$^{1}$ and 
  M~V~Vorontsov$^{1}$
  }
  \address{
    $^{1}$ Peter the Great St. Petersburg Polytechnic University, 195251, Saint~Petersburg, Russian~Federation
    \\  
    $^{2}$ Ioffe Institute,  194021, Saint~Petersburg, Russian~Federation
  }
  \ead{goglichidze@gmail.com}
  \begin{abstract} 
    We consider the pulsar rotation assuming that the neutron star consists of crust component (which rotation is observed) and two core components. 
    One of the core components contains pinned superfluid which can, for some reasons, suddenly inject small fraction of stored angular momentum in it. In the framework of this simple model the star can demonstrate glitch-like events together with long period precession (with period $1-10^{4}$ years). 
  \end{abstract}

  %                 P0                tau_0   B_surf
  %B1828-11   0.405043321630        1.07e+05 4.99e+12

  \section{Introduction}     
    Some radio pulsars demonstrate periodic variations in pulse timing and beam shape.
    The most favorable explanation to these phenomena is the precession of the neutron star with periods $T_p \sim 10^2$ days \cite{AshtonJonesPrix2016,KerrHobbsJohnstonShannon2016}. %StairsLyneShemar2000
    There are pulsar characteristics and behavior features which can be interpreted as a manifestation of precession with much larger periods $T_p\sim10^2-10^4$ years \cite{ArzamasskiyEtAl2015,BiryukovBeskinKarpov2012}.
    Another 
    pulsar timing feature 
%    type of timing features 
%     type of timing irregularities 
    is pulsar glitches which are the sudden 
    increase of pulsar frequency
%     increase of pulsar angular velocity $\Omega$ 
    with subsequent smooth recovery.
    Standard model relates these events with avalanche-like unpinning of superfluid neutron vortices pinned to the nuclei in the inner crust or proton flux tubes in the neutron star core. However, as it was first pointed out by Shaham \cite{Shaham1977} and later confirmed by more detailed researches \cite{SedrakianWassermanCordes1999,Link2006}, the pinning of neutron superfluid should dramatically decrease the period of precession making it very far from observed values. 
    We propose a simple phenomenological model which allows the same pulsar to precess with a long period and demonstrate glitch-like 
    behavior.
%     timing features.
    
  \section{Three-components model}
    Let us assume that a neutron star consists of three dynamically distinguished components. For the sake of simplicity all components are supposed to rotate as rigid bodies.    

    \textbf{The c-component} is the outer component.
%     including NS crust. 
    It rotates with angular velocity $\vec{\Omega}_c = \vec{\Omega}$:
    \begin{equation}
      \label{eq:dt_Mc}
      d_t\vec{M}_{c} =  \vec{K}_\mathrm{ext} + \vec{N}_{rc} + \vec{N}_{gc},
    \end{equation}
    where $\vec{M}_{c} = I_{c} \vec{\Omega} + I_{c} \hat{\epsilon} \vec{\Omega}$ is the $c$-component angular momentum, $I_{c}$ is one of its principal moments of inertia.  Tensor $\hat{\epsilon}$ is the effective oblateness tensor. 
    It describes the total departure of c-component mass distribution and electromagnetic field energy distribution from spherical symmetry.
%     It describes the nonsphericity of  and star electromagnetic field inertia distributions (which manifests as anomalous torque \cite{GBT2015}).
%     Here, we implicitly assume that the magnetic field anchored to the c-component.
    It is assumed that $\epsilon_{\alpha\beta} \ll 1$.
%     and we suppose that tensor $\hat{\epsilon}$ is the only source of neutron star precession.
%     Vector $\vec{K}_\mathrm{ext} \propto \Omega^3 $ is the external torque which is the cause of NS secular rotation evolution.
    Vector $\vec{K}_\mathrm{ext}$ is the external electromagnetic torque. Note that so-called anomalous torque is taken into account by tensor $\hat{\epsilon}$ \cite{GBT2015}.
    Therefore, torque $\vec{K}_\mathrm{ext}$ contains only the terms $\propto \Omega^3$.
%     which is the cause of NS secular rotation evolution.
    
    \textbf{The g-component} is an inner component. It consists of charged matter characterized by moment of inertia $I_g$ and rotating with angular velocity $\vec{\Omega}_g$. It also contains neutron superfluid with angular momentum $\vec{L}_g$ pinned to the charged matter:  
    \begin{equation}
      \label{eq:dt_Mg}
%       d_t{\vec{M}}_{g} = I_{g} d_t \vec{\Omega}_{g} + \left[ \vec{\Omega}_{g} \times \vec{L}_{g} \right] = - \vec{N}_{gc} - \vec{N}_{gr},
      d_t{\vec{M}}_{g}  = \vec{N}_{cg} + \vec{N}_{rg}, \ \ \ d_t \vec{L}_g = \vec{\Omega}_g\times\vec{L}_g,
    \end{equation}
    where $\vec{M}_{g} = I_g\vec{\Omega}_g + \vec{L}_g$ is the total angular momentum of the g-component.
%     , $I_{g}$ is the moment of inertia of normal matter.
%     , $\vec{N}_{g}$  is the torque due to interaction with other components.
    
    \textbf{The r-component} is the other inner component. It rotates with angular velocity $\vec{\Omega}_{r}$:
    \begin{equation}
      \label{eq:dt_Mr}
%       d_t\vec{M}_{r} = I_{r} d_t\vec{\Omega}_{r} = - \vec{N}_{rc} + \vec{N}_{gr},
      d_t\vec{M}_{r}  =  \vec{N}_{cr} + \vec{N}_{gr}
    \end{equation}
    where $\vec{M}_{r}= I_{r} \vec{\Omega}_{r}$ is the angular momentum of the r-component, $I_{r}$ is its moment of inertia.
%     , $\vec{N}_{r}$  is torque due to interaction with other components. 
    The components interact with each other by torques $\vec{N}_{ij}$, where $i,j = c,g,r$ and, obviously, $\vec{N}_{ij}=-\vec{N}_{ji}$.

%     Generally the components interact with each other by the torque $\vec{N}_{ij}$ which by assumption has the form
%     \begin{equation}
%       \label{eq:N_ij}
%       \vec{N}_{ij} = - I_{j} \left( \alpha_{ij} \vec{\mu}^{||}_{ij} + \beta_{ij} \vec{\mu}^{\perp}_{ij}  +\gamma_{ij} [ \vec{e}_{\Omega} \times \vec{\mu}^{\perp}_{ij} ]\right).
%     \end{equation}
%     Here, $\alpha_{ij}, \beta_{ij}, \gamma_{ij}$ are some interaction constants,  
%     $\vec{\mu}_{ij} = \vec{\Omega}_{j} - \vec{\Omega}_{i}$, 
%     $i,j = (c,g,r)$,   
%     $\vec{e}_{\Omega} =\vec{\Omega}/\Omega$,
%     and we have introduced notations $A^{||} = (\eO \cdot \vec{A})$ and $\vec{A}^{\perp} = \vec{A} - \eO\, ( \eO \cdot \vec{A})$ applied to arbitrary vector $\vec{A}$. It is obvious that $\vec{N}_{ij} = - \vec{N}_{ji}$.
    
%     First let us note that, 
    If $\vec{K}_\mathrm{ext} = 0$ and $\hat{\epsilon}=0$, system of equations \eqref{eq:dt_Mc}-\eqref{eq:dt_Mr} has a simple equilibrium solution: $\vec{\Omega}_g = \vec{\Omega}_r = \vec{\Omega} = \mbox{const}$, $\vec{L}_g = L_g \vec{e}_{\Omega}$, where $\vec{e}_\Omega = \vec{\Omega}/\Omega$.
    Further we will consider only small corrections to this equilibrium state.  
    Introducing notations $V^{||} = (\eO \cdot \vec{V})$ and $\vec{V}^{\perp} = \vec{V} - \eO\, ( \eO \cdot \vec{V})$ applied to arbitrary vector $\vec{V}$ and
    holding only the linear in  $\vec{\mu}_{ij} = \vec{\Omega}_{j} - \vec{\Omega}_{i}$ and $\vec{L}_{g}^{\perp}$ terms one can rewrite system of equations  \eqref{eq:dt_Mc}-\eqref{eq:dt_Mr} in the following form: \\
    \begin{minipage}{0.43\textwidth}
      \begin{align}
        &\dot{\Omega}  =  R_{gc}^{||} + R_{rc}^{||} + S^{||},
        \label{dOmega_linear_pp_eqn}
        \\
        &\dot{\mu}_{cg}^{||}  = 
          R_{cg}^{||} + R_{rg}^{||}
        - R_{gc}^{||} - R_{rc}^{||} - S^{||},
        \label{dmucg_linear_pp_eqn} 
        \\
        &\dot{\mu}_{cr}^{||}  = 
        R_{cr}^{||} + R_{gr}^{||} - R_{gc}^{||} - R_{rc}^{||} - S^{||},
        \label{dmucr_linear_pp_eqn}
        \\
        &\dot{\omega}_{g}^{||}  =  0,
        \label{dLg_linear_pp_eqn}  
        \\
        &\Omega \dot{\vec{e}}_{\Omega}  =  
        \vec{R}_{gc}^{\perp} + \vec{R}_{rc}^{\perp} + \vec{S}^{\perp},
        \label{dOmega_linear_perp_eqn}        
      \end{align}
    \end{minipage}
    \begin{minipage}{0.56\textwidth}
      \begin{align}
        \label{dmucg_linear_perp_eqn} 
        &\dot{\vec{\mu}}_{cg}^{\perp}  =  
          (\omega_{g}^{||} - \Omega) \, 
          [ \vec{e}_{\Omega} \times \vec{\mu}_{cg}^{\perp} ]
        - [\vec{\Omega} \times \vec{\omega}_{g}^{\perp}]
        + \\
        \nonumber
         & \ \ \ \ + \vec{R}_{cg}^{\perp} + \vec{R}_{rg}^{\perp} 
        - \vec{R}_{gc}^{\perp} - \vec{R}_{rc}^{\perp} - \vec{S}^{\perp},
        \\
        \label{dmucr_linear_perp_eqn}        
        &\dot{\vec{\mu}}_{cr}^{\perp}  =  
        - [\vec{\Omega} \times \vec{\mu}_{cr}^{\perp} ] + \\ 
        & \ \ \ \ + \vec{R}_{cr}^{\perp} + \vec{R}_{gr}^{\perp} 
        - \vec{R}_{rc}^{\perp} - \vec{R}_{gc}^{\perp} - \vec{S}^{\perp},
        \nonumber
        \\
        &\dot{\vec{\omega}}_{g}^{\perp}  = 
        - \frac{\omega_{g}^{||}}{\Omega} \,
          \left(
             \vec{R}_{gc}^{\perp} + \vec{R}_{rc}^{\perp} + \vec{S}^{\perp} 
           + [ \vec{\Omega} \times \vec{\mu}_{cg}^{\perp}]
          \right),
        \label{dLg_linear_perp_eqn}          
      \end{align}
    \end{minipage} \\ \\ \\
    where 
          $\vec{R}_{ij} = \vec{N}_{ij} / I_{j}$,
          $\vec{S} = \vec{K}_\mathrm{ext} / I_{c} - \vec{\Omega}\times \hat{\epsilon}\vec{\Omega}$,
          $\vec{\omega}_{g} = \vec{L}_{g} / I_{g}$,
          and the point denotes the time derivative in the c-component frame of reference.
%    \hl{We have also neglected small term} $I_c \hat{\epsilon}\dot{\vec{\Omega}}$ \hl{in equation} \eqref{eq:dt_Mc}.
%    Without loss of generality
    We assume that vectors $\vec{R}_{ij}$ linearized in $\mu_{ij}$ can be represented as     
    \begin{equation}
       \label{eq:R_ij}
       \vec{R}_{ij} = - \left( \alpha_{ij} {\mu}^{||}_{ij}\vec{e}_\Omega + \beta_{ij} \vec{\mu}^{\perp}_{ij}  +\gamma_{ij} [ \vec{e}_{\Omega} \times \vec{\mu}^{\perp}_{ij} ]\right),
    \end{equation}
    where $\alpha_{ij}, \beta_{ij}$ and $\gamma_{ij}$ are the phenomenologically introduced interaction constants. 

%    \hl{From the obvious equality} $\vec{N}_{ij}=-\vec{N}_{ji}$ \hl{one can obtain that}  $\alpha_{ij} =   (I_i/I_j)  \alpha_{ji}$, $\beta_{ij} =   (I_i/I_j)  \beta_{ji}$, $\gamma_{ij} =   (I_i/I_j)  \gamma_{ji}$.
%    $\vec{R}_{ij} = - (I_i/I_j) \vec{R}_{ji}$.
  \section{Quasi-stationary rotation}
    Let us first assume that glitches do not occur.
    In this case, if the largest internal relaxation time-scale $\tau_{rel} = {\mathrm{max}}\left( 1 /\alpha_{ij}, 1/\beta_{ij}, 1/\gamma_{ij} \right)$ is much smaller than the period of precession $T_p$, and assumption \eqref{eq:R_ij} is valid,
%     and time-scale of secular evolution 
    the neutron star will rotate in quasi-stationary regime. 
    It means that,
%      if the age of neutron star larger than $\tau_{rel}$, 
    the departures of internal components rotation from c-component rotation are
%      with good accuracy 
    determined by instant values of $\vec{\Omega}$ and $\dot{\vec{\Omega}}$ and do not depend on prehistory (see detail in \cite{BGT2013}). In other words, one can neglect time derivative terms in equations \eqref{dmucg_linear_pp_eqn}, \eqref{dmucr_linear_pp_eqn}, \eqref{dmucg_linear_perp_eqn}-\eqref{dLg_linear_perp_eqn} because of their quadratic smallness.
    Solving equations \eqref{dmucg_linear_pp_eqn}, \eqref{dmucr_linear_pp_eqn}, \eqref{dmucr_linear_perp_eqn} and \eqref{dLg_linear_perp_eqn} for $\mu_{cr}^{||}$, $\mu_{cg}^{||}$ $\vec{\mu}_{cr}^{\perp}$ and $\vec{\mu}_{cg}^{\perp}$, and substituting the obtained expressions into equations \eqref{dOmega_linear_pp_eqn} and \eqref{dOmega_linear_perp_eqn}, one can obtain
    \begin{equation}
      \label{eq:retard}
      \tilde{I}_\mathrm{tot}\dot{\Omega} = K_\mathrm{ext}^{||},
    \end{equation}
    \begin{equation}
      \label{eq:precess}
      I_c \Omega \dot{\vec{e}}_{\Omega} =\frac{(1+\Gamma)(\vec{K}_\mathrm{ext}^\perp-I_c\vec{\Omega}\times\hat{\epsilon}\vec{\Omega})+B\vec{e}_\Omega\times(\vec{K}_\mathrm{ext}^\perp-I_c\vec{\Omega}\times\hat{\epsilon}\vec{\Omega})}{(1+\Gamma)^2+B^2},
    \end{equation}
    where $\tilde{I}_\mathrm{tot} = I_c + I_g + I_r$
    and 
%     coefficients $B$ and $\Gamma$ 
    in the weak interaction limit  ($\alpha_{ij}, \beta_{ij},\gamma_{ij} \ll \Omega$) equals to
    \begin{equation}
      \label{eq:precess_coeffs_weak_limit}
      B \approx \frac{\beta_{gc} + \beta_{rc}}{\Omega} \ll 1, \ \ \ \ \Gamma \approx \frac{\gamma_{gc} + \gamma_{rc}}{\Omega} \ll 1.
    \end{equation}
    
    One can note that according to equation \eqref{eq:retard} the neutron star is braked as if it is a solid body with moment of inertia $\tilde{I}_\mathrm{tot}$ (which does not include the moment of inertia of pinned superfluid).
%     and the retardation rate does not depend on the exact mechanism of angular momentum transfer from the core into the crust. 
    It is a general feature of quasi-stationary approximation \cite[section 2.3]{BGT2014}.    
    
    Equation \eqref{eq:precess} describes the change in the orientation of angular velocity vector $\vec{\Omega}$ relative to the c-component.
%     neutron star precession and its damping (or growing). 
    The first and the third terms on the right-hand side arise due to action of external torque $\vec{K}_\mathrm{ext}$, the second term makes vector $\vec{\Omega}$ precess about a star principal axis and the fourth term describes the damping of the precession due to internal energy dissipation. 
%     Formally the limiting case of ``rigid'' star corresponds to $B=0$, $\Gamma = I_\mathrm{tot}/I_c-1$, where $I_\mathrm{tot}$ is the total moment of inertia. 
%     If components interaction is [[[rather]]] weak ($\beta_{ij},\gamma_{ij} \ll \Omega$), coefficients \eqref{eq:precess_coeffs} takes the form:
    %
    Using equation \eqref{eq:precess},
%     and the smallness of coefficients $B$ and $\Gamma$, 
    one can estimate precession period and precession damping time-scale as
    \begin{equation}
      \label{eq:T_p-tau_p}
      T_p \sim \frac{2\pi}{\epsilon\Omega}, \ \ \
      %       \approx 10^{12}\frac{P_{sec}}{\epsilon_{12}}\mbox{ sec } = 1.15\times10^7 \frac{P_{sec}}{\epsilon_{12}}\mbox{ days }= 3.85\times 10^5 \frac{P_{sec}}{\epsilon_{12}} \mbox{ months } = 3.17\times10^4 \frac{P_{sec}}{\epsilon_{12}} \mbox{ years },
      \tau_d \sim \frac{2\pi}{\epsilon\Omega B} \gg T_p,   
    \end{equation}
    %     In this case the precession period can be estimated as 
%     \begin{equation}
%       T_p \sim \frac{2\pi}{\epsilon\Omega}\approx 10^{12}\frac{P_{sec}}{\epsilon_{12}}\mbox{ sec } = 1.15\times10^7 \frac{P_{sec}}{\epsilon_{12}}\mbox{ days }= 3.85\times 10^5 \frac{P_{sec}}{\epsilon_{12}} \mbox{ months } = 3.17\times10^4 \frac{P_{sec}}{\epsilon_{12}} \mbox{ years },
%     \end{equation}
    where $\epsilon$ is characteristic oblateness. 
    The external torque $\vec{K}_\mathrm{ext}$ tends to minimize the energy loss caused by electromagnetic radiation. 
    It leads to increase or decrease of precession amplitude depending on the orientation of star dipole moment relative to the principal axes \cite{Goldreich1970}.
    One can see that expression for precession period $T_p$ does not contain $L_g$. This seems to contradict to Shaham who has obtained that $T_p$ equals rather to  $2\pi(\tilde{I}_\mathrm{tot}/L_g)$ \cite{Shaham1977}. The reason for this discrepancy is that we allow the g-component (which containing the pinned superfluid) to move relative to the c-component which rotation is observed. It is required that $\beta_{cg}\ll\Omega$ to ensure the validity of expressions \eqref{eq:T_p-tau_p}. The Shaham's result can be reproduced by passing to limit  $\beta_{cg}\rightarrow \infty$ in equations \eqref{dmucg_linear_perp_eqn}-\eqref{dLg_linear_perp_eqn} so that $\vec{\mu}_{cg}^{\perp}\rightarrow 0$.  
%     In this case, after the quasistatic regime is reached, ......
%     Keeping in mind inequalities \eqref{eq:precess_coeffs_weak_limit} one can estimate the characteristic evolution time-scale as
%     \begin{equation}
%       \tau_{ev} \sim \frac{I_c\Omega}{K_\mathrm{ext}} \approx 1.3\times10^8 P_{sec}^2 \frac{I_{c45}}{B_{12}^{2}} \mbox{ years},
%     \end{equation}   
%     where $I_{c45} = I_{c}/10^{45}$ g cm$^2$, $B_{12}$ is the induction of magnetic field measured at the magnetic poles in units of $10^{12}$ G. 
%     Precession is also damped by energy dissipation  in the internal layers of the star. The rate of dissipation depends on coefficient $B$. Therefore, remember again inequalities \eqref{eq:precess_coeffs_weak_limit} one can estimate the precession damping time-scale as
%     \begin{equation}
%       \tau_d \sim \frac{P}{B\epsilon} \gg T_p.
%     \end{equation}

  \section{Glitch-like event}
    Since superfluid is pinned, the lag between rotations of superfluid neutrons and charged part of g-component increases during the retardation of star rotation. When critical lag is reached the superfluid unpins and then repins such that some small fraction of its angular momentum  is transferred from superfluid into the charged fraction. 
        \begin{wrapfigure}{r}{0.4\linewidth}
          \center
          \vspace{-1cm}
          \includegraphics[width=0.6\linewidth,trim= 0.4cm 1.05cm 0.6cm 0.cm, clip = true]{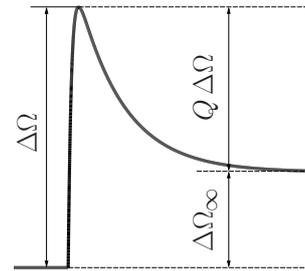}
          \caption{\label{fig:glitch}The sketch of the glitch-like behavior of $\Omega$.}        
    \end{wrapfigure}               
    In present work we do not specify the physical mechanism of glitch triggering. However, we assume that the glitch occurs and then relaxes at time-scales much less the the time-scales of quasi-stationary evolution. It allows us to neglect the effects of external torque and star oblateness considering these processes.
    
    The solution of equations \eqref{dOmega_linear_pp_eqn}-\eqref{dmucr_linear_pp_eqn} for $\Omega$ with the following initial conditions: $\Omega = \Omega_0$, $\mu_{cr}^{||} = 0$, $\mu_{cg}^{||} = \Delta L_g/I_g$ and zeroth ${S}^{||}$ has the form
    \begin{equation}
      \label{eq:Omega_glitch}
      \Omega(t) = \Omega_0 + \Delta\Omega
      \left( 1 - e^{-p_{+}t} - Q (1-e^{-p_{-}t}) \right),
    \end{equation}
    where 
    $\Delta\Omega = {\Delta \Omega_{\infty}}/{(1-Q)}$,
    $\Delta\Omega_{\infty} = { \Delta L_{g}}/{\tilde{I}_\mathrm{tot}}$,
    $Q = (\tilde{I}_\mathrm{tot} \alpha_{cg} - I_{c} p_{+} )/( \tilde{I}_\mathrm{tot} \alpha_{cg} - I_{c} p_{-} )$ and
    the coefficients $p_{+}$ and $p_{-}$ ($p_{+} > p_{-}$)
    are the roots of equation
    \begin{eqnarray}
      p^{2} & - &
      \left( \alpha_{cg} + \alpha_{rg} + \alpha_{cr} + \alpha_{gr} 
           + \alpha_{gc} + \alpha_{rc}
      \right) \, p 
      +
      \nonumber
      \\
       & + &
        \left( \alpha_{gc} + \alpha_{rg} + \alpha_{cg} \right) \cdot
        \left( \alpha_{cr} + \alpha_{gr} + \alpha_{rc} \right)
      + \left( \alpha_{rc} - \alpha_{rg} \right) \cdot
        \left( \alpha_{gr} - \alpha_{gc} \right)
      = 0.
      \label{p_def} 
    \end{eqnarray}
    The meaning of parameters $\Delta\Omega$, $\Omega_\infty$ and $Q$ becomes clear if one looks at the sketch of the solution in figure \ref{fig:glitch}.
    We suppose that the interaction between the c- and  g-components is the strongest one ($\alpha_{cg} \gg  \alpha_{rc}, \alpha_{rg}$). In this case,
    \begin{equation}
      \label{eq:p+p-Q}
      p_{+} \approx \left( 1 + \frac{I_{g}}{I_{c}} \right) \alpha_{cg}, \  \ \ p_{-} \approx \frac{\tilde{I}_\mathrm{tot}}{I_{c} + I_{g}} \left( \alpha_{cr} + \alpha_{gr} \right)
      \mbox{\ \ and \ \ }
      Q \approx \frac{I_{r}}{\tilde{I}_\mathrm{tot}}.
    \end{equation}
    If one wants to relate solution \eqref{eq:Omega_glitch} with observed pulsar glitches, then $1/p_{+}$ and $1/p_{-}$ should be interpreted as glitch growth ($ < 30$ s \cite{DodsonLewisMcCulloch2007}) and relaxation ($1-10^2$ days \cite{LyneShemarSmith2000}) times respectively.       
    One can see that the glitch growth rate is proportional to $\alpha_{cg}$
%       depends on the interaction between c- and g-components 
    while the subsequent relaxation is governed by the r-component. 
    Thus, on the one hand, the interaction between the c-  and g-components should be strong enough to ensure rapid spin-up.
%       during the glitch. 
    But simultaneously it should  be not too much strong, namely $\beta_{cg} \ll \Omega$, to long period precession could exist 
    It is natural to assume that $\alpha_{cg} \sim \beta_{cg}$.    
%       Otherwise, the precession damping time-scale $\tau_d$ becomes smaller than the period of precession $T_p$.
%       Recovery fraction $Q$ is approximately proportional to $I_r$. Therefore, 
  
%       As before we consider a neutron star which does not feel the action of external torque ($\vec{S} = 0$). Let us assume that before $t=0$ the star was in equilibrium state. At time t=0 
          
  \section{Discussion}    
    Let us speculate a little about possible nature of formally introduced components.
    The c-component can be associated with neutron star crust and a part of core charged particles which is strongly coupled with the crust ($I_c\sim10^{-2}-10^{-1}I_\mathrm{ns}$). 
    The role of g-component can be performed by tangles of closed flux tubes which could be formed after protons became superconductive from chaotic small-scale magnetic field.
%     presenting in  the core of young neutron star. 
     Alternatively it can be a torus composed of closed flux tubes.
%      formed from the toroidal magnetic field 
     \cite{GugercinougluAlpar2014}. If the characteristic cross-section $S_\mathrm{tor}$ of the region occupied by the toroidal field is of the order of 1 km$^{2}$, then $I_g\sim \rho_p S_\mathrm{tor} r_\mathrm{ns}^3 \sim 10^{-3}I_{ns}$, where $\rho_p$ is the proton mass density.
    Some part of superfluid neutron vortices presented in the core should be pinned to the flux tubes. On the one hand, this interaction prevents the tangles or torus from collapse. On the other hand, when the critical rotational lag is reached, the vortices unpin from the flux tubes triggering the glitch.
    Since the flux tubes are closed in the core, the g-component, being magnetically decoupled, can rotate with angular velocity different from $\vec{\Omega}$ \cite{GlampedakisLasky2015}.
    The r-component can be represented by the part of core neutron superfluid
    which is not pinned to the g-component and a part of normal matter coexisting with it and weakly coupled with c-component ($I_r\sim I_\mathrm{ns}$).
    %\hl{Strictly speaking, it would be more consistent to treat superfluid and normal fractions of r-component as  two distinguished components because of their weak interaction. It would complicate the calculations and probably could lead to more complex glitch relaxation but qualitatively the model remains the same.}
    %
    The interaction coefficients values which are needed to make the model able to explain the observed glitches time-scales can be estimated with expressions \eqref{eq:p+p-Q}.
    The weakest side of the presented simple model is the value of recovery fraction $Q$. According to \eqref{eq:p+p-Q} it is of the order of unity. It is not so bad for young pulsars but mature ones demonstrate rather recovery fractions several orders less than unity \cite{LyneShemarSmith2000}. However, we believe that it is a consequence of the oversimplifying of the model and further researches will allow us to avoid this discrepancy. 
      
  \section*{References}
    \bibliography{iopart-num,mn-jour,paper}   

\end{document}